\documentclass[useAMS,usenatbib]{mn2e}

\usepackage{epsfig}
\newcommand{\hr}{$^{\rm h}$}
\newcommand{\mn}{$^{\rm ^m}$}

\newcommand{\hi}{H{\sc i}}
\newcommand{\amin}{$`$}
\newcommand{\kms}{\mbox{km s$^{-1}$}}

\title[The Origin of Large-scale \hi\ structures in the Magellanic Bridge]{The Origin of Large-scale \hi\ structures in the Magellanic Bridge}
\author[Erik Muller,  Kenji Bekki]{Erik Muller$^{1}$\thanks{E-mail:
erik.muller@csiro.au} and Kenji Bekki$^{2}$\thanks{E-mail:bekki@phys.unsw.edu.au}\\
$^{1}$Australia Telescope National Facility, CSIRO, P.O. Box 76, Epping, NSW, 1710, Australia\\
$^{2}$School of Physics, University of New South Wales, Sydney, NSW, 2052, Australia}
\begin{document}

\date{}

\pagerange{\pageref{firstpage}--\pageref{lastpage}} \pubyear{2007}

\maketitle

\label{firstpage}

\begin{abstract}
We investigate the formation of a number of key large-scale \hi\
features in the ISM of the Magellanic Bridge using dissipationless
numerical simulation techniques. This study comprises the first direct comparison between detailed \hi\ maps of the Bridge and numerical simulations. We confirm that the SMC forms two tidal filaments: a near arm, which forms the connection between the SMC and LMC, and a counter-arm. We show that the \hi\ of the most dense part of the Bridge can become arranged into a bimodal
configuration, and that the formation of a 'loop' of \hi, located
off the North-Eastern edge of the SMC can be reproduced simply as a projection of the counter-arm, and without invoking localised energy-deposition processes such as SNe or
stellar winds.
\end{abstract}


\section{Introduction}
The formation and evolution of the Interstellar Medium (ISM) is
affected by processes which cause the deposition of energy across
a complete range of spatial scales:  galactic collisions and
interactions deposit energy into the ISM over scales which are
comparable to the dimensions of the galaxies themselves
\citep[e.g.][]{berentzen,goldman,gsf94}; while accumulated energy
from Stellar wind, Supernova events (SNes) or Gamma-Ray Bursts
(GRBs), can deposit approximately 10$^{51}$ ergs per event into a
relatively confined volume \citep[e.g.][]{bloom,lozinskaya} and
reorganise the ISM on kiloparsec scales \citep[e.g.][]{dbw2000}.
Energy injected into a system via these and other mechanisms can
subsequently propagate down through smaller spatial scales
according to a power-law \citep[e.g.][]{goldman}. In order to
obtain a full understanding of the processes active in shaping the
ISM, it is clear that studies of the large-scale structure are
necessary \emph{in addition} to smaller-scale analyses.

The Magellanic Bridge (MB) is the epitome of tidal features and
its origin as such is well established as the product of an
interaction between the \hi-rich Small and Large Magellanic Clouds
(SMC, LMC), some 200 Myr ago \citep[e.g.][]{gsf94,yn2004}. At
50-60 kpc \citep[e.g.][]{abraha}, the Magellanic System represents the closest
interacting system to our Galaxy.

The ISM in the MB and SMC is turbulent, and conforms to a featureless,
Kolmogorov power spectrum \citep{muller2004,stani_smc} from kpc scales, down
to the limit of current radio observations of $\sim$30 pc. Even
so, the arrangement of structure of the ISM in the MB is not
completely homogeneous at the larger scales: \cite{muller2003}
\citep[See also][]{wayte}
identified three statistically interesting features which dominate
the overall structure of the densest parts of the MB. These
manifest as; 1) a distinct and separable high-velocity \hi\
component, which exists only at the more northerly declinations,
and appears shifted in velocity from the brightest \hi\ component
by $\sim$35\kms (See 'Counter arm' feature in Figure~\ref{fig:fig2} E2); 2) a large (R$\sim$1.3 kpc) loop,
located off the north-east edge of the body of the SMC (see Figure~\ref{fig:fig1} A,B) and;
3) a bimodal arrangement of the brightest (i.e. most dense) parts of the MB (See Figure~\ref{fig:fig2} E1). 
\citet{muller2004} were able to show that the higher velocity
component (feature \#1, from above) is morphologically distinct
from the southern, denser component; the range of velocity
modifications (i.e. spurious modifications to the 3-D power distribution by
a turbulent component) to the ISM within the northern part are
significantly smaller than for the southern part. These authors
argue that the distinction in turbulence and power-structure
between the north-south regions is consistent with numerical
simulations by \cite{gsf94}, which predict that the two
components represent two arms emanating from the SMC. In this
case, the northern part is the projection of an almost radially
extending arm which does not form a contiguous link between the
SMC and LMC. The more southern component comprises matter drawn
out from the SMC body following the SMC-LMC interaction. It
remains to understand the evolution of features \#2 and \#3.

Although the formation of the Magellanic Stream and the
evolution of the LMC and the SMC have been already  investigated
by previous numerical simulations on tidal interaction between the
Clouds and the Galaxy \citep[e.g.][]{gsf94,bekki05,connors06}, the
formation of the MB has not been extensively investigated on a
detailed level. Here, we compare our observational results on \hi\
structure and kinematics with the corresponding simulations and
discuss the tidal interaction model in the context of the
observations. We will also discuss alternative formation
mechanisms and processes and we will see that although  the
processes that shape the ISM of the Magellanic Bridge are not yet
fully understood to a fine, detailed level, we can at least begin
to understand the mechanisms which are responsible for the
development of the larger-scale structures in this unique
filament.

\section{The \hi\ dataset}
Being the primary constituent of the ISM, \hi\ is the most useful
probe of its bulk and turbulent motions. We make extensive use of
a lower-resolution dataset of the entire Magellanic system,
obtained by \cite{bruns} using the Parkes\footnote{The Australia
Telescope Compact Array and Parkes Telescopes are part of the
Australia Telescope, which is funded by the Commonwealth of
Australia for operation as a National Facility managed by CSIRO}
telescope. These data have a sensitivity of 0.05 K and a spatial
and velocity resolution of 16$'$ and 1 \kms. Other details of the
measurements and analysis of the lower-resolution dataset are
covered in \cite{bruns}. We also make a frequent reference to a
high-resolution observations of the western MB only (1.5\hr$<$RA$<\sim$3\hr), using the ATCA and
Parkes telescopes, which have a sensitivity to 0.8 K per
$\sim$1.6 \kms\ channel, and a spatial resolution of $\sim$98$"$. A
description of the observations and reduction process of the
high-resolution dataset can be found in \cite{muller2003}. The
features which form the focus of this simulation study were
identified in work by \cite{muller2003}, using the high-resolution
dataset.

\subsection{Loop feature and Bimodal structure}
The \hi\ loop manifests as an enormous, localised deficiency of
material in the line of sight; subtending an ellipse with axes
$\sim$1.6 and 1.0 kpc. The loop is marked in Figure~\ref{fig:fig1}A,
and appears to be located in \hi\ that is 
shifted by $\sim$40\kms\ relative to the bulk of the Bridge: approximately 190.7-231.9 \kms\ (LSR). Based on the mean column density around the loop
\citep[$\sim$5$\times$10$^{20}$cm$^{-1}$;][]{muller2003}, the loop
appears to represent an \hi\ deficiency (i.e. the mass of material
that appears to be missing) of $\sim$2$\times$10$^{7} {\rm M}_{\odot}$.

Previous proposals regarding the origins of the loop include
speculation its position corresponds to that of the second
SMC-LMC Lagrange point \citep{wayte}, however the concept is not
subjected to a quantitative test.\\

The brightest part of the \hi\ in the Bridge (roughly bounded by Dec
$-$72$^{\circ}$30\amin\ to $-$73$^{\circ}$30\amin\ and 1\hr\ 30\mn\ to
2\hr\ 30\mn)  has been shown to be organised into two 'sheets'
\citep{muller2003}, approximately parallel in velocity and separated
by $\sim$30 \kms. The feature can be seen in Figure~\ref{fig:fig2}E1,
where the bimodal arrangement appears to originate in the eastern edge
of the SMC and extends eastward. It is clearly a dominant feature
in the MB, and involves $\sim$8$\times$10$^7$${\rm M}_{\odot}$ of
\hi; approximately 80\% of the total \hi\ mass of the Western MB.  
We may gauge the significance of this structure by temorarily assuming it
is a kinematic process: $\sim$8$\times$10$^7$${\rm M}_{\odot}$ of \hi\ expanding at $\sim$30 \kms\ requires approximately $1/2 \sum MV^2 =
9\times10^{52}$ erg. This feature is clearly indicative of a
large-scale and energetic process and is worthy of attention when
attempting to understand the formation of the Bridge.

\section{Numerical Simulations}\label{sec:n_sims}
We investigate dynamical evolution of the Clouds from 0.8 Gyr
ago ($T=-0.8$ Gyr)  to the present ($T=0$) by using numerical
simulations in which the SMC gas particles are modelled in a self-gravitating
disky system, and the LMC is a test particle. The key epoch for the MB
formation is 0.2 Gyr when the SMC passed its pericenter distance with
respect to the LMC \citep[e.g.][]{gn96}.

In developing the simulations, we search for the models which most closely reproduce
(1) the current locations of the Clouds and the MB, (2) an
apparently contiguous \hi\ filament clearly seen in the sky, and
(3) the \emph{total} mass of $\sim 10^8 {\rm M}_{\odot}$ in  the
MB. It is necessary to explore a very wide parameter space for
orbits, masses, and disk inclinations of the LMC and the SMC and
we therefore investigate the formation processes of the MB based
on {\it dissipationless simulations} in the present study. Our
future studies based on full-blown chemodynamical simulations such
as \cite{bekki05} will discuss the importance of hydrodynamics and
star formation in the MB formation. A discussion of fundamental
methods and techniques of numerical simulations on the evolution
of the Clouds is given in previous papers \citep{bekki05} and
these will not be re-addressed here.

\subsection{Initial Conditions}
The model SMC is composed of stellar disk, and collisionless "gas" disk, embedded in a massive dark halo having a 'NFW' profile \citep{nfw}.  Typically for gas-rich systems, the \hi\ diameters are much larger than the optical disk
\citep[e.g.][]{bvw94} and we configure the simulations where the radius of the SMC gas disk ($R_{\rm g}$) is twice as large as its stellar disks ($R_{\rm s}$). We use masses of the LMC and the SMC that are consistent with observations by \cite{marel} and \cite{ss97}.

The Galactic gravitational potential ${\Phi}_{\rm G}$ is represented by ${\Phi}_{\rm G}=-{V_{0}}^2 {\rm ln} (r)$, where $V_{0}$ and $r$ are the constant rotational velocity (220 km s$^{-1}$ in this study) and the distance from the Galactic centre. The orbits of the SMC and LMC are bound.

We use the same coordinate system  $(X,Y,Z)$ (in units of kpc) as
those used in  \cite{gn96} and \cite{bekki05}. The adopted current
positions are $(-1.0,-40.8,-26.8)$  for the LMC and
$(13.6,-34.3,-39.8)$ for the SMC and the adopted current
Galactocentric radial velocity of the LMC (SMC) is 80 (7) km
s$^{-1}$. Current velocities of the LMC and the SMC in the
Galactic ($U$, $V$, $W$) coordinate are assumed to be
(-5,-225,194) and  (40,-185,171) in units of km s$^{-1}$,
respectively.

The initial spin of a SMC disk in a model is specified by two
angles, $\theta$ and $\phi$, where $\theta$ is the angle between
the $Z$-axis and the vector of the angular momentum of a disk and
$\phi$ is the azimuthal angle measured from $X$-axis to the
projection of the angular momentum vector of a disk onto the $X-Y$
plane.

Although we investigate a large number of models with different
$R_{\rm g}/R_{\rm s}$, $\theta$ and $\phi$, we show only the
results of the two best-performing models; Model 1 and Model 2.
The fundamental parameters of these models are summarised in
Table~\ref{tab:models}.

\begin{table}
\centerline{
\begin{tabular}{lcc}\\\hline\hline
&Model 1&Model 2\\\hline
$R_{\rm g}/R_{\rm s}$&2&2\\
$M_{\rm L}$&$2.0 \times 10^{10}{\rm M}_{\odot}$&$2.0 \times 10^{10}{\rm M}_{\odot}$\\
$M_{\rm s}$&$1.5 \times 10^{9} {\rm M}_{\odot}$&$1.5 \times 10^{9}{\rm M}_{\odot}$\\
$\theta$&-45$^{\circ}$&-45$^{\circ}$\\
$\phi$&270$^{\circ}$&330$^{\circ}$\\\hline
\end{tabular}
}
\caption{\label{tab:models}Summary of System parameters for two most successful simulations of the MB.}
\end{table}

\subsection{Simulation results}
\begin{figure}
\centerline{\resizebox{7.9cm}{!}{\includegraphics{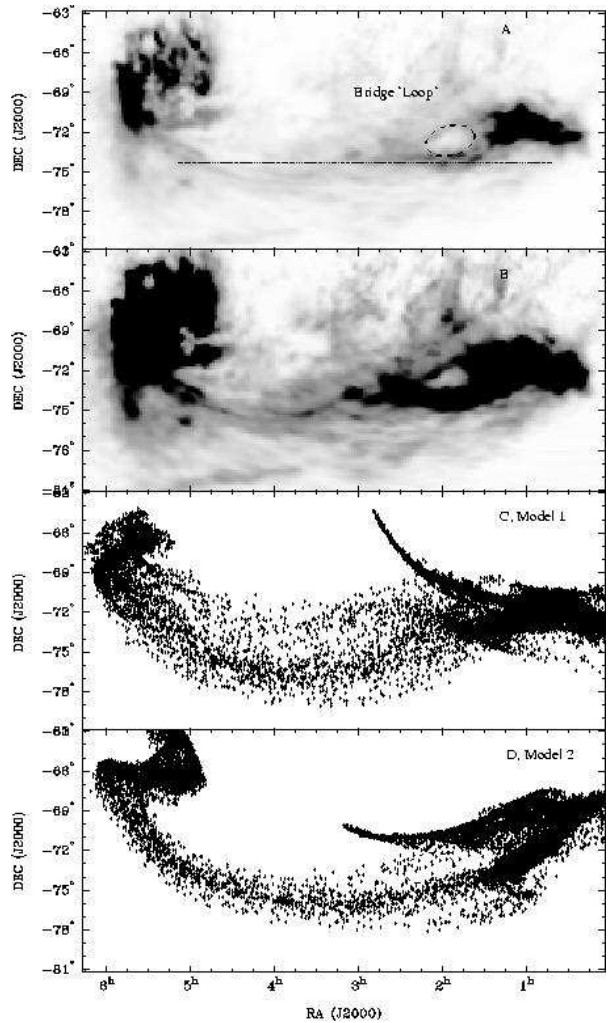}}}
\caption{\label{fig:fig1}\emph{(Top)} Panel A (RA,Dec) shows the
observed distribution of integrated \hi\ with the Magellanic
Bridge \citep{bruns}. Panel A is strongly contrasted in panel B to highlight the loop feature.  the horizontal line indicates $-$73$^{\circ}$, for reference in Figure\ref{fig:fig2}. Panels C and D show the projected
distribution of modelled gas-particles for Models 1 and 2 respectively. }
\end{figure}

\begin{figure*}
\centerline{\resizebox{15cm}{!}{\includegraphics{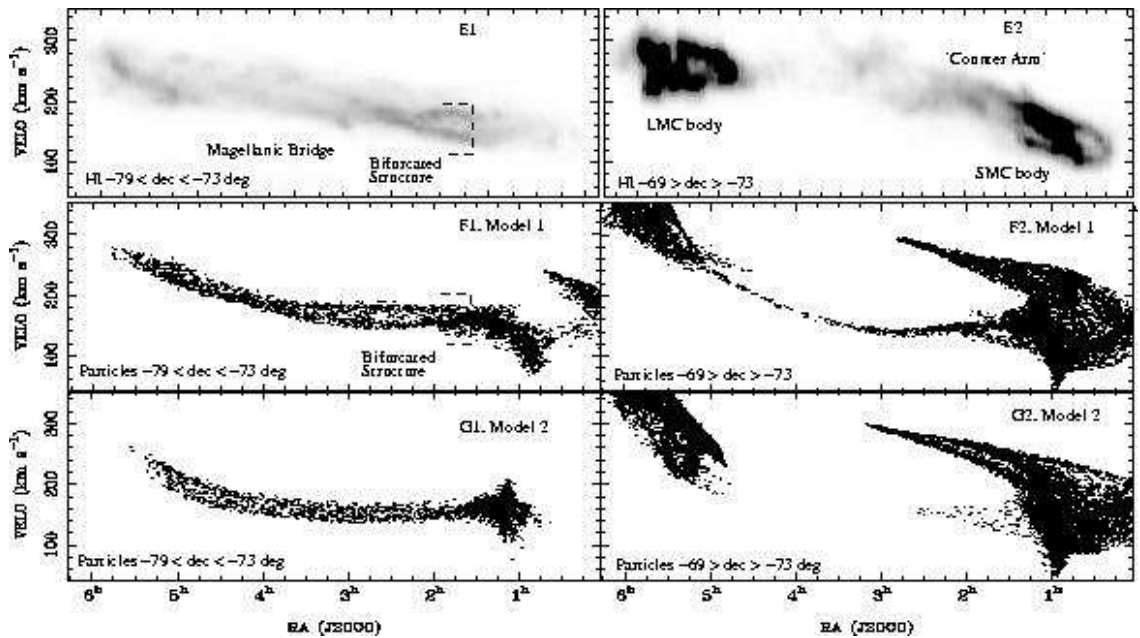}}}
\caption{\label{fig:fig2}\emph{(Top row)} Panels E1 and E2 show
the RA-velocity projection of neutral hydrogen for $-$79$^{\circ}< $Dec
$<-$73$^{\circ}$ and  $-$69$^{\circ}> $Dec $> -$73$^{\circ}$. Panels F1 and F2 show
the distribution of simulated gas-particles for Model 1, while panels
G1 and G2 show the distribution of gas-particles for Model 2.}
\end{figure*}

From a qualitative study of Figure~\ref{fig:fig1}, we see that the
large-scale and general filamentary arrangement of the numerical
simulations are consistent with observations: the dense
'Bridge' filament is well reproduced, as well as the central body of the
SMC.

The velocity projections in Figure~\ref{fig:fig2} show that the
bimodal properties of the densest part of the Bridge can easily be
duplicated by Model 1. The two distinct velocity ranges represent
similar orbital 'families' which are present in the Bridge itself.
The simulations are able to duplicate the observed velocity
separation of $\sim$40\kms\, as well as the approximate spatial
extent of the bimodal arrangement for ${\rm 2^{h}0^{m}} \le \alpha
\le {\rm 3^{h}0^{m}}$ and 150  km s$^{-1}$ $\le V_{\rm h} \le$
200  km s$^{-1}$. A significant quantity of material is drawn into
the more northerly declinations and out into a velocity range
which is different to that of the 'Bridge' itself.

Both models 1\&2 reproduce the SMC as having two filaments: a lower-declination, low-velocity
and nearer arm (forming the Bridge proper), and a higher-declination, high-velocity
counter-arm component that has more radial extension (See also Figure~\ref{fig:fig4}). These
results are completely consistent with earlier numerical
simulations by \cite{gsf94}, and with a quantitative analysis of
the turbulent structure of the \hi\ around the SMC by
\cite{muller2004}.

\begin{figure}
\centerline{
\resizebox{8cm}{!}{\rotatebox{-90}{\includegraphics{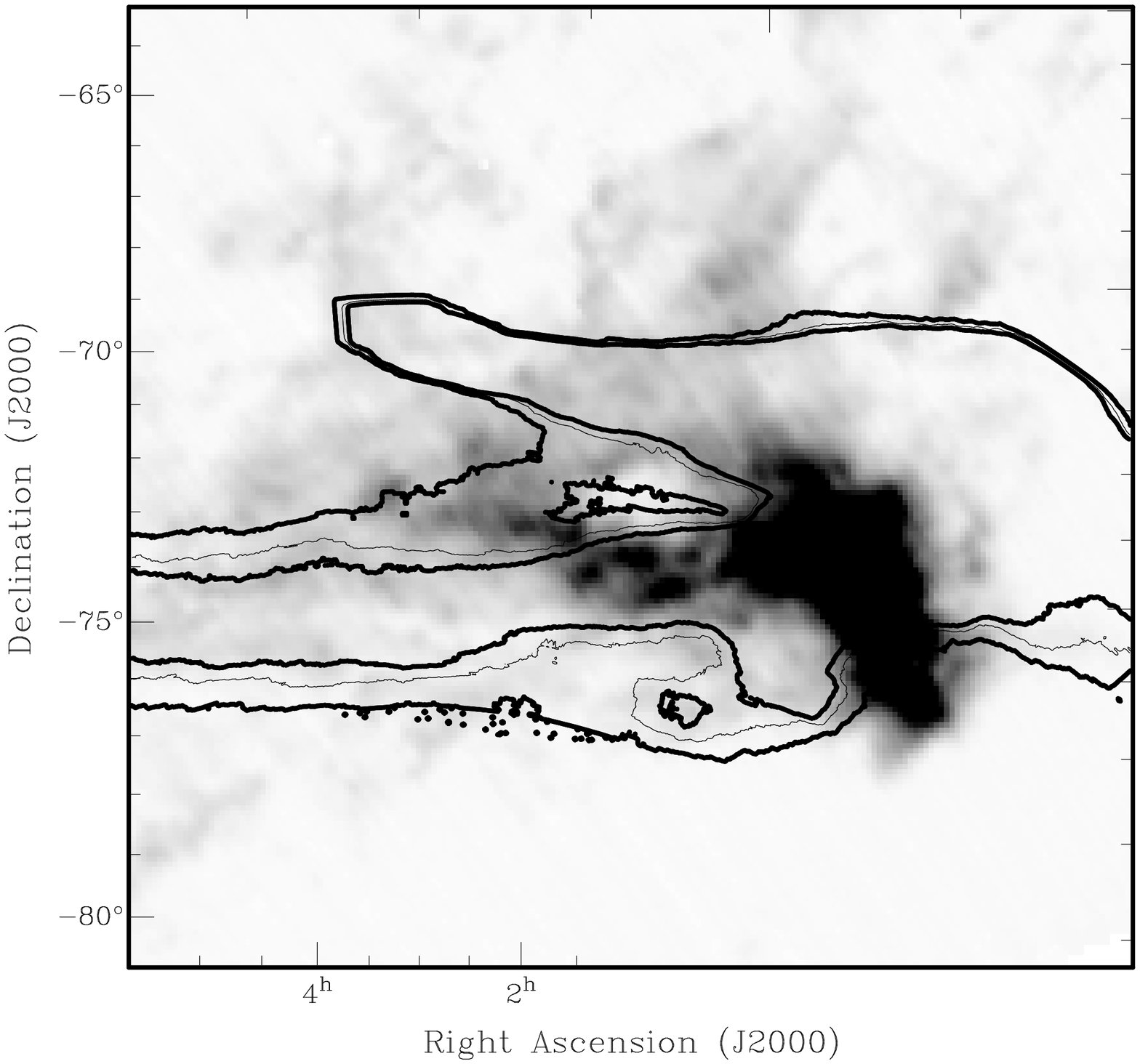}}}}
\caption{\label{fig:fig3} \emph{Greyscale:} Magnified subregion of
the \hi\ brightness distribution around the SMC \citep{bruns}.
\emph{Contours:} smoothed contours from particles predicted by
Model 2. The 'loop' is central to the figure, and its formation
has been clearly reproduced in this simulation. Note that the
contour levels are set very low ($<$1\%), and so the simulated SMC
appears to have a large extent.}
\end{figure}

Finally, we see in Figure~\ref{fig:fig3} that model 2 can
convincingly reproduce the observed 'loop' off the north-eastern
corner of the SMC. \cite{muller2003} report that the loop is found within a
contiguous velocity range. A study of their Figures 3 and 4 reveals that the bulk of the loop is consistent with the velocity-shifted northern component, and that the bottom of the loop is delimited by the lower-velocity and brighter southern component.

\begin{figure*}
\centerline{ \resizebox{15cm}{!}{\includegraphics{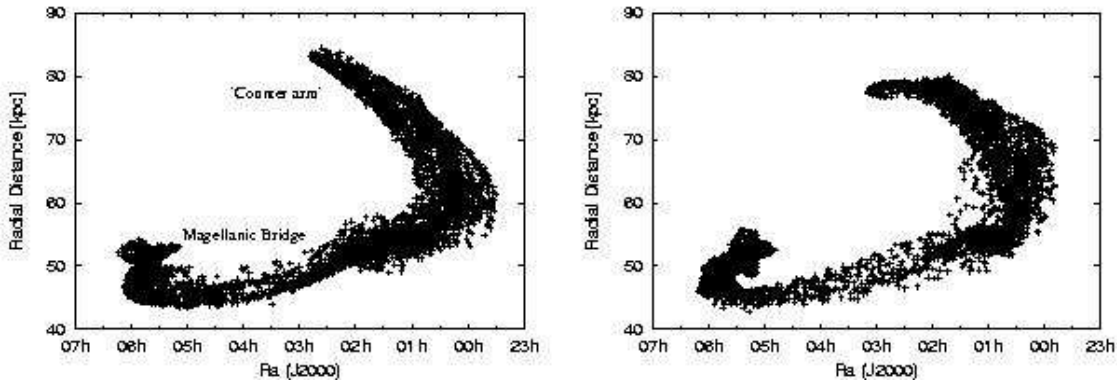}}}
\caption{\label{fig:fig4} RA and radial position of gas-particles from
Models 1 ({\emph left}) and 2 ({\emph right}). Both models show
the existence of a radially-extending arm. Every 16th point is
plotted to reduce crowding.}
\end{figure*}

We show in Figure~\ref{fig:fig4} the position-radius projection of
the simulations. We see that in both cases, two filaments emanate from the forming SMC. The actual Magellanic Bridge may be regarded as the
more nearby component, whereas the tidal counterpart to the Bridge
extends more radially. Importantly, both models predict an
extremely large line-of-sight depth for the SMC (including the Bridge, SMC and counter-arm) consistent with some previous numerical simulation results \citep[e.g.][]{gsf94}; and also a line of sight depth through the Bridge which is consistent with the $\sim$5
kpc line-of-sight measured between adjacent stellar clusters by
\cite{demers} in the Magellanic Bridge.

Although neither of the two models can reproduce the observations
fully self-consistently, we find that the three key large-scale
features in the Magellanic Bridge are produced by these two
models. This suggests that the scenario where the Bridge is the
result of a tidal interaction between the Clouds and the Galaxy
for the last 0.2 Gyr is essentially important in explaining
fundamental properties of large-scale organisation of the ISM in
the MB.

The lack of exact reproduction at this stage may be due to
insufficiently complex simulations, which exclude ISM feedback
processes. We note that other large-scale numerical simulations,
such as those by \cite{gsf94} are also unable to clearly reproduce the loop
feature. Future and more sophisticated models with gas dynamics
and star formation will confirm whether the tidal interaction
model can explain both  the two kinematical properties  (i.e., the
bimodal kinematics and the velocity offset) and the presence of
the giant \hi\ loop in a self-consistent manner.

\section{Alternative loop formation scenarios}
Given the preliminary nature of these results in predicting the
formation of the Bridge 'loop', it is appropriate to explore
alternative processes which also may develop similar structures such
as \hi\ 'shells'. Processes such
as the stellar-wind, SNe and HVC impacts are commonly-cited in the
literature as shell-formation mechanisms.

Using the canonical formulations to estimate the total input
energy by \cite{weaver} (shell evolution powered by stellar winds) along with data of the basic characteristics of the observed loop
(R=1.3 kpc), and based on its observed velocity position in the
'high velocity' component which has a limiting velocity
dispersion of $\sim$30 \kms, we estimate an energy requirements for the shell expansion: E$_{Weav}$= 53.5 log ergs. Energies of this
magnitude are equivalent to that provided by stellar winds, or SNe
produced from $\sim$100 O-type stars \citep{chevalier}. In any other larger system,
an association of a few hundred stars would be unremarkable.
However, OB associations numbering more than $\sim$10 are uncommon
in the Magellanic Bridge \citep[][;Bica, priv.  comm. 2001]{bica} and is not
clear how such a relatively well populated association may come to
be at the observed location. For this reason, formation of the
loop by stellar winds (or SNe) is considered implausible.

Works by \cite{tb88} and \cite{tt86} have shown through a variety
of two dimensional numerical simulations, how an \hi\ cloud
infalling into a stratified gas layer can generate an expanding
shell-like structure. From \citet{tt86}, we can relate the
expansion energy of the hole to the kinetic energy of a
hypothetical infalling cloud, via the cloud's density and radius:

 n[cm$^{-3}$]=9.78$\times$ 10$^{-43}\frac{E_{kin}}{R_c^3V_c^2}$

Using the kinetic energy estimated previously, $E_{kin}$ (erg), we
can probe the ranges of HVC properties that are capable of
generating a hole having the same observed parameters of the
Bridge loop. We find that after limiting the density of the
candidate infalling HVC to 0.2$< \rho >$5 cm $^{-3}$, according to
the range of HVC densities estimated by \cite{brunsphd}, a 0.6
$\times$10$^6$ ${\rm M}_{\odot}$ cloud need only move at a
velocity of $\sim$-350 \kms\ in the LSR frame to attain the
estimated kinetic energy to create the observed loop. Such a
velocity is not unusual for many of the known HVC population
\citep[e.g.][]{wop2002,putman2002}. Recently \cite{bekki06} have
found  that massive sub-halos can create kpc-scale giant HI holes
such as those seen in the Western Magellanic Bridge. These results
imply that if the MB interacted with low-mass subhalos, which are
predicted by a hierarchical clustering scenario to be ubiquitous
in the Galactic halo region, giant HI holes can be formed. We plan
to investigate this possibility in our forthcoming papers.

{}
\bsp

\label{lastpage}

\end{document}